\documentclass[a4paper]{article}

\usepackage[english]{babel}
\usepackage[utf8x]{inputenc}
\usepackage[T1]{fontenc}

\usepackage[a4paper,top=3cm,bottom=2cm,left=3cm,right=3cm,marginparwidth=1.75cm]{geometry}

\usepackage{amsmath}
\usepackage{graphicx}
\usepackage[colorinlistoftodos]{todonotes}
\usepackage[colorlinks=true, allcolors=blue]{hyperref}

\title{Resource allocation in C-V2X: A review}
\author{Tahmid Zaman Tahi : \space The George Washington University   }

\begin{document}
\maketitle

\section*{Abstract}
C-V2X is a rapidly evolving technology that opened countless opportunities for IoV applications. In this paper, we explore the existing research on resource allocation for both LTE-V2X and NR-V2X and take a peek into the future trends for research in this field.  

\section{Introduction}

\space \space			 Cellular Vehicle-to-Everything (C-V2X) is a cutting-edge wireless communication technology that enables seamless connectivity and information exchange among vehicles, infrastructure, networks, and pedestrians. As a vital component of Intelligent Transportation Systems (ITS), C-V2X is designed to support a wide range of applications aimed at enhancing traffic efficiency, improving road safety, reducing accident rates, and facilitating the development of autonomous and connected vehicles.
\newline \space C-V2X technology is built upon the Long-Term Evolution (LTE) and 5G New Radio (NR) standards, leveraging the robustness, reliability, and scalability of cellular networks. It encompasses two distinct communication modes: (1) direct communication, which includes Vehicle-to-Vehicle (V2V), Vehicle-to-Infrastructure (V2I), and Vehicle-to-Pedestrian (V2P) communication, and (2) network-based communication, which involves Vehicle-to-Network (V2N) communication.
\newline \space Resource allocation is a critical challenge in the design and operation of C-V2X systems, as it is responsible for determining the optimal distribution of communication resources among users, ensuring efficient utilization and fair sharing. In C-V2X, resource allocation is complicated by factors such as highly dynamic network topologies, diverse quality of service (QoS) requirements, and spectrum scarcity. Therefore, it is essential to explore and analyze various resource allocation strategies and techniques that can effectively address these challenges.
\newline \space This review paper provides a comprehensive overview of the recent progress in resource allocation for C-V2X communications. As C-V2X technology evolves, it is expected to play a crucial role in transforming the transportation landscape, paving the way for smarter, safer, and more efficient transportation systems.

\section{Resource Allocation in LTE-V2X}
\subsection{Basic sidelink resource allocation modes in LTE-V2X}
LTE-V2X communication has two resource allocation (RA) modes: Mode 3 and Mode 4.

Mode 3 functions under cellular coverage, with the eNB (Evolved NodeB) allocating resources to vehicles. There are two RA options in this mode: 1. Dynamic allocation and 2. Semi-persistent scheduling (SPS). Dynamic allocation necessitates that each vehicle request subchannels from the eNB for each CAM packet transmission, while SPS reserves subchannels for vehicles' periodic transmissions. SPS is more appropriate for V2V services due to the recurring nature of basic safety messages.

Mode 4, standardized by 3GPP for autonomous radio resource selection, employs a sensing-based SPS algorithm. Vehicles independently choose radio resources by sensing the channel and selecting a resource from a pool of candidate resources. The algorithm consists of three main stages: (1) reserving candidate resources within a selection window (SW), (2) generating a list L1 of resources, and (3) creating a list of resources L2 with the lowest RSSI values. At that point, the vehicle arbitrarily selects a concluding resource within the L2 list. [1]

LTE-V2X mode 4 also accommodates resource allocation based on geographical zone positions. However, packet collision issues may arise due to channel load, so Channel Occupancy Ratio (CR) and Channel Busy Ratio (CBR) are utilized to assess channel congestion.

To avoid conflicts between vehicles functioning in Modes 3 and 4, resources are set up distinctly for each mode. In Release-15, the Mode 3 SCI field signifies the SPS booking duration for the allocated resource, which allows Mode 4 vehicles to disregard resources used by Mode 3 vehicles while executing the SPS algorithm. [5]
 
\subsection{Retransmission schemes}
\space \space \space \space \space \space \space \space \space C-V2X in LTE uses sidelink channel introduced in 3GPP release 12 for D2D communications. But sidelink in LTE does have any feedback channels, which gave rise to researchers proposing retransmission schemes. 

Juyeop Kim et al. introduced an index coded retransmission scheme to enhance the performance of V2X communication, specifically focusing on mitigating redundant retransmissions in V2X direct communication. This scheme effectively incorporates the index coding process with the retransmission process, reducing the consumption of radio resources due to redundant retransmissions. By XOR-ing the redundant blocks of the two initial transmissions and transmitting them after the two initial systematic blocks have been transmitted, the use of resource blocks is considerably decreased, making the approach more cost-effective [2].

Donglin Wang et al. proposed a retransmission scheme that performs one blind retransmission after the initial transmission, irrespective of the success of the initial transmission. This scheme involves transmitting the initial message and then conducting a blind retransmission with different traffic schemes to test the system's reliability. Redundant transmissions are executed randomly after a time gap, and their success is dependent on the channel's condition. However, as the redundant retransmissions are not contingent on the success of the initial transmission, this scheme consumes more resources than the index-coded retransmission process, making it less resource-efficient [3].

Md. Imrul Hassan et al. proposed and evaluated a retransmission scheme based on the performance of IEEE 802.11p for dedicated short-range communication (DSRC). In their paper, two retransmission-based schemes are put forward to enhance the reliability and efficiency of broadcasts within a contention-based MAC in DSRC [4]. 
 
\subsection{New research in RA for LTE-V2X}
\subsubsection{Mode 3 for LTE-V2X}
 Abanto-Leon et al. developed a resource allocation (RA) solution to tackle intracluster and interclusters interference problems that occur when vehicles are grouped by their geographical location. However, the vehicle clustering technique faces challenges, primarily due to the high mobility of vehicles and cluster stability. The authors do not offer comprehensive information about their clustering approach, and a suboptimal clustering strategy may negatively impact the RA scheme's effectiveness [6].

Other research has concentrated on RA schemes rooted in geographical positions and distance-based resource reuse methods, such as those presented by Cecchini et al. [7], Fritzsche and Festag [8], and Sempere-García [9]. While these algorithms exhibit promising performance, they might experience degradation due to inaccurate vehicle positioning. Additionally, they do not account for the varying Quality of Service (QoS) requirements of different V2X applications, including safety, traffic management, and entertainment.

Allouch et al. developed the PEARL algorithm, which caters to the QoS requirements of diverse V2X applications by dividing traffic into safety and nonsafety classifications. The algorithm reserves resources for safety traffic and shares the remaining resources between safety and nonsafety traffic. In congested traffic conditions, the PEARL algorithm implements a priority-based RA scheme, prioritizing safety messages over nonsafety messages [10]. The PEARL-RM algorithm extends the PEARL algorithm by incorporating a resource reuse mechanism (RM) to optimize resource allocation for both safety and nonsafety traffic [11].

Khabaz et al. proposed the MIRD algorithm, which emphasizes RA for essential safety traffic by clustering vehicles based on their geographical locations and directions. The algorithm assigns orthogonal resource sets to each cluster and uses a unique resource reuse approach between clusters, determined by the intercentroids distance. This strategy ensures effective resource allocation, particularly in areas with high vehicle density, and avoids blocking transmitters [12].

\subsubsection{Mode 4 for LTE-V2X} In recent years, research on mode 4 in vehicle-to-everything (V2X) communications has been categorized into three main areas. These categories concentrate on evaluating the performance of the semi-persistent scheduling (SPS) algorithm, enhancing the SPS algorithm, and suggesting new alternatives to the SPS algorithm.

The first category, encompassing the majority of research works, aims to assess and optimize the SPS algorithm's performance. In this category, R. Molina-Masegosa and J. Gozalvez conducted the first study evaluating the SPS algorithm under realistic traffic scenarios [13]. Nabil et al. illustrated the impact of the resource reservation interval (RRI) on packet data rate performance [14]. Molina-Masegosa et al. revealed that increasing the probability of reselection could enhance the SPS algorithm's performance under certain conditions [15]. Gonzalez-Martín employed analytical models to assess the SPS algorithm's performance [16], while Bartoletti et al. investigated the influence of cooperative awareness message (CAM) packet generation on mode 4 performance [17].

The second category of research efforts focuses on adapting the SPS algorithm to enhance the performance of mode 4. Bonjourn et al. proposed a counter learning and reselection (CLR) mechanism to tackle collision issues [18]. Jeon and Kim introduced a new resource reservation solution aimed at minimizing packet collisions [19]. They suggested a short-term sensing-based resource selection (STS-RS) scheme [20], while Abanto-Leon et al. put forth a new nonlinear power averaging method [21]. Bazzi et al. addressed the problem of wireless blind spots (WBS) in their study [22], and Molina-Masegosa and Gozalvez dealt with resource allocation for packets of varying sizes [23].

Lastly, the third category seeks to present alternative approaches to mode 4 that do not rely on incremental enhancements to the SPS algorithm. Sabeeh et al. introduced the estimation and reservation resource allocation (ERRA) algorithm [24], while Yang et al. presented the predictive assessment of resource usage (PRESS) algorithm [25]. Zhao et al. recommended an autonomous cluster-based resource selection scheme [26], and another RA approach was proposed in [27] to alleviate the hidden node problem. Sahin and Boban concentrated on resource allocation in delimited out-of-coverage areas (DOCA) by suggesting an algorithm that enhances reliability for V2V communication [28].

In conclusion, research concerning mode 4 in V2X communications covers a broad range of approaches, from assessing and optimizing the SPS algorithm to proposing new alternatives for enhanced performance. These studies significantly contribute to the understanding and advancement of mode 4 in V2X communication systems.

\section{Resource Allocation in NR-V2X}
\subsection{Basic sidelink resource allocation in NR-V2X}
Both LTE-V2X and NR-V2X support two resource allocation (RA) modes: mode 1 (under cellular coverage) and mode 2 (out-of-cellular-coverage). Allocating the resource pool between these modes can lead to more efficient radio resource utilization; however, it may also cause collisions between vehicles operating in distinct modes. To address this issue, mode 1 vehicles notify mode 2 vehicles of their future resource allocations via the SCI field.

Mode 1 uses dynamic grant (DG) scheduling and configured grant (CG) scheduling, both of which involve communication with the gNB. Mode 2 enables vehicles to independently choose resources configured by the gNB or pre-set within the vehicles. To accommodate new use cases in NR-V2X, 3GPP members have contemplated improvements to the sensing-based SPS algorithm employed in LTE-V2X mode 4. They propose adjusting the sensing and resource selection procedures to cater to both periodic and non-periodic traffic.

The two channel sensing techniques which were proposed include long-term sensing (employed in LTE-V2X mode 4) and short-term sensing (which involves observing the medium prior to transmission). Long-term sensing is well-suited for periodic traffic, whereas short-term sensing is better for non-periodic traffic. Combining both approaches is also a possible option. Given that NR-V2X modes were established by 3GPP rather recently, additional research work is required to examine these modes' functionality and verify their efficacy in practical scenarios.

\subsection{New research in RA for NR-V2X}
\subsubsection{Mode 1 for NR-V2X}
Xiaoqin et al. introduced a new RA algorithm for mode 1, aimed at minimizing energy consumption when transmitting CSI from vehicles to the gNB [29]. The authors formulated the sidelink radio RA issue as an MBINP to optimize system throughput. However, they did not account for the effects of high vehicle mobility on their proposed algorithm.

Abbas et al. presented a two-tier RA strategy to reduce latency, enhance throughput, and ensure reliability for V2V and V2N communications. They explored a scenario where a relay vehicle is needed to forward signals between the gNB and a transmitting vehicle. The authors designed an effective RA algorithm to prevent interference and satisfy QoS requirements for both V2V and V2N communications, with simulations demonstrating improved throughput and reduced latency [30].

Gao et al. proposed a DNN-based RA algorithm for mode 1, primarily focusing on maximizing system throughput by optimally allocating transmit power to vehicles [31].
\subsubsection{Mode 2 for NR-V2X}
The review paper emphasizes recent research on sensing-based SPS algorithms for NR-V2X's autonomous mode 2. Ali et al. assessed the influence of different parameters on system performance [32], while Romeo et al. concentrated on the transmission reliability of Decentralized Environmental Notification Messages (DENM) [33]. Their findings revealed that short-term sensing with numerous repetitions enhanced the DENM delivery ratio but impacted reliability when the DENM message load grew.

Yoon and Kim proposed a stochastic reservation scheme for aperiodic traffic, which predicted the next transmission time for resource reservation [34]. This proposal showed high Packet Reception Rate (PRR) values. Meanwhile, Yi et al. proposed a novel sensing-based SPS method to efficiently allocate resources in congested intersection areas using RSUs as central controllers [35].

Researchers have also explored the effects of NR flexible numerologies on the performance of autonomous modes. Ali et al. and Campolo et al. showed that higher SubCarrier Spacing (SCS) improved reliability and PRR, while reducing Update Delay [36] [37]. However, these studies did not consider delay spread and vehicle speeds as critical parameters.

Machine Learning (ML) techniques, such as deep reinforcement learning, have been proposed for new RA algorithms, with the aim of minimizing interferences and meeting V2V communication requirements. [38]

\section{Future research directions} In this review paper, we explore the forthcoming research avenues associated with resource allocation (RA) in Cellular Vehicle-to-Everything (C-V2X) technology., focusing on enhancements and emerging technologies that can improve vehicular communication systems. One of the primary enhancements under discussion for 3GPP Release 17 includes reducing power consumption in NR sidelink mode 2, which could potentially lower the energy consumption of user equipment (UE) devices. Additionally, inter-UE coordination is being considered to improve communication reliability and reduce packet collisions [39].

Emerging technologies such as machine learning (ML), mobile-edge computing (MEC), and network slicing (NS) present promising opportunities for advancing RA in C-V2X. Machine learning methods can be utilized to effectively allocate resources to vehicles using historical data, addressing the unique characteristics of vehicular networks [40] [41]. Mobile-edge computing, or more specifically vehicular edge computing (VEC), can be utilized to process and transmit large volumes of data generated by advanced onboard sensors in connected vehicles, with radio RA playing a crucial role alongside computational RA [42] [43]. Network slicing, which allows for the creation of multiple virtual networks on a shared physical infrastructure, can facilitate the efficient allocation of resources to meet the diverse requirements of various V2X applications.
 
\section{Conclusion} 
In this paper, the existing research works in mode 3 and mode 4 for LTE-V2X and mode 1 and mode 2 for NR-V2X are presented. While there has been significant research on LTE-V2X mode 4, mode 3 has received less attention. Several areas still require further exploration, such as the functioning of future submode 2(d), resource allocation between modes 1 and 2, mode-switching, and managing NR sidelink resources via eNB or LTE sidelink resources through gNB. To overcome these challenges and leverage emerging technologies, researchers must continue to develop innovative resource allocation algorithms to improve 5G-V2X communication and fulfill the demands of ultra-reliable, low-latency communication services.

\newpage
\bibliographystyle{alpha}

\bibliography{sample}
\space [1]  M. Gonzalez-Martín, M. Sepulcre, R. Molina-Masegosa, and
J. Gozalvez, “Analytical models of the performance of C-V2X mode 4
vehicular communications,” IEEE Trans. Veh. Technol., vol. 68, no. 2,
pp. 1155–1166, Feb. 2019.

[2] J. Kim, D. K. Sung, and S. H. Kim, "Index Coded Retransmission Scheme for V2X Direct Communication," in IEEE Wireless Communications Letters, vol. 7, no. 5, pp. 824-827, October 2018. doi: 10.1109/LWC.2018.2835159

[3] D. Wang, X. Wu, R. Wang, Z. Li, X. Lin, and H. Zhu, "A Retransmission Scheme Based on Channel Load Prediction for V2X Communication," in IEEE Access, vol. 7, pp. 109542-109554, 2019. doi: 10.1109/ACCESS.2019.2932460

[4] M. I. Hassan, S. S. Kanhere, S. Chawla, and M. Tariq, "Retransmission-Based Broadcast in Vehicular Ad Hoc Networks," in IEEE Transactions on Vehicular Technology, vol. 66, no. 7, pp. 6351-6365, July 2017. doi: 10.1109/TVT.2016.2629547    

[5] “Technical specification group radio access network, overall description
of radio access network (RAN) aspects for vehicle-to-everything (V2X)
based on LTE and NR,” 3GPP, Sophia Antipolis, France, Rep. TR
37.985, Jun. 2020.

[6] L. F. Abanto-Leon, A. Koppelaar, and S. H. de Groot, “Subchannel
allocation for vehicle-to-vehicle broadcast communications in mode-3,”
in Proc. IEEE Wireless Commun. Netw. Conf. (WCNC), 2018, pp. 1–6.

[7] G. Cecchini, A. Bazzi, B. M. Masini, and A. Zanella, “Localizationbased resource selection schemes for network-controlled LTE-V2V,” in
Proc. Int. Symp. Wireless Commun. Syst. (ISWCS), 2017, pp. 396–401.

[8] G. Cecchini, A. Bazzi, M. Menarini, B. M. Masini, and A. Zanella,
“Maximum reuse distance scheduling for cellular-V2X sidelink mode
3,” in Proc. IEEE Globecom Workshops (GC Wkshps), 2018, pp. 1–6.

[9] R. Fritzsche and A. Festag, “Location-based scheduling for cellular
V2V systems in highway scenarios,” in Proc. IEEE 87th Veh. Technol.
Conf. (VTC Spring), 2018, pp. 1–5.

[10] M. Allouch, S. Kallel, A. Soua, and S. Tohme, “PEARL: A novel
priority and guaranteed-based resource allocation approach for V2V
communications in LTE-V mode 3,” in Proc. 8th Int. Conf. Wireless
Netw. Mobile Commun. (WINCOM), 2020, pp. 1–6.

[11] M. Allouch, S. Khemiri-Kallel, A. Soua, and S. Tohme, “A priority and
guarantee-based resource allocation with reuse mechanism in LTE-V
mode 3,” in Proc. Wireless Days (WD), 2021, pp. 1–5.

[12] S. Khabaz, T.-M.-T. Nguyen, G. Pujolle, and P. B. Velloso, “A new
clustering-based radio resource allocation scheme for C-V2X,” in Proc.
12th Wireless Days Conf. (WD), Paris, France, Jun. 2021, pp. 1–8.

[13] R. Molina-Masegosa and J. Gozalvez, “System level evaluation of LTEV2V mode 4 communications and its distributed scheduling,” in Proc.
IEEE 85th Veh. Technol. Conf. (VTC Spring), 2017, pp. 1–5.

[14] A. Nabil, K. Kaur, C. Dietrich, and V. Marojevic, “Performance analysis of sensing-based semi-persistent scheduling in C-V2X networks,”
in Proc. IEEE 88th Veh. Technol. Conf. (VTC-Fall), 2018, pp. 1–5.

[15] R. Molina-Masegosa, J. Gozalvez, and M. Sepulcre, “Configuration of
the C-V2X mode 4 sidelink PC5 interface for vehicular communication,” in Proc. 14th Int. Conf. Mobile Ad-Hoc Sens. Netw. (MSN), 2018,
pp. 43–48

[16] M. Gonzalez-Martín, M. Sepulcre, R. Molina-Masegosa, and
J. Gozalvez, “Analytical models of the performance of C-V2X mode 4
vehicular communications,” IEEE Trans. Veh. Technol., vol. 68, no. 2,
pp. 1155–1166, Feb. 2019.

[17] S. Bartoletti, B. M. Masini, V. Martinez, I. Sarris, and A. Bazzi,
“Impact of the generation interval on the performance of sidelink
C-V2X autonomous mode,” IEEE Access, vol. 9, pp. 35121–35135,
2021.

[18] N. Bonjorn, F. Foukalas, F. Cañellas, and P. Pop, “Cooperative resource
allocation and scheduling for 5G eV2X services,” IEEE Access, vol. 7,
pp. 58212–58220, 2019.

[19] Y. Jeon and H. Kim, “An explicit reservation-augmented resource allocation scheme for C-V2X sidelink mode 4,” IEEE Access, vol. 8,
pp. 147241–147255, 2020.

[20] X. He, J. Lv, J. Zhao, X. Hou, and T. Luo, “Design and analysis
of a short-term sensing-based resource selection scheme for C-V2X
networks,” IEEE Internet Things J., vol. 7, no. 11, pp. 11209–11222,
Nov. 2020.

[21] L. F. Abanto-Leon, A. Koppelaar, and S. H. de Groot, “Enhanced
C-V2X mode-4 subchannel selection,” in Proc. IEEE 88th Veh.
Technol. Conf. (VTC-Fall), 2018, pp. 1–5.

[22] A. Bazzi, C. Campolo, A. Molinaro, A. O. Berthet, B. M. Masini, and
A. Zanella, “On wireless blind spots in the C-V2X sidelink,” IEEE
Trans. Veh. Technol., vol. 69, no. 8, pp. 9239–9243, Aug. 2020.

[23] R. Molina-Masegosa and J. Gozalvez, “LTE-V for sidelink 5G V2X
vehicular communications: A new 5G technology for short-range
vehicle-to-everything communications,” IEEE Veh. Technol. Mag.,
vol. 12, no. 4, pp. 30–39, Dec. 2017.

[24] S. Sabeeh, P. Sroka, and K. Wesołowski, “Estimation and reservation for autonomous resource selection in C-V2X mode 4,” in Proc.
IEEE 30th Annu. Int. Symp. Personal Indoor Mobile Radio Commun.
(PIMRC), 2019, pp. 1–6.

[25] J. M. Yang, H. Yoon, S. Hwang, and S. Bahk, “PRESS: Predictive
assessment of resource usage for C-V2V mode 4,” in Proc. IEEE
Wireless Commun. Netw. Conf. (WCNC), 2021, pp. 1–6.

[26] J. Zhao et al., “Cluster-based resource selection scheme for 5G V2X,”
in Proc. IEEE 89th Veh. Technol. Conf. (VTC-Spring), 2019, pp. 1–5.

[27] R. Molina-Masegosa, M. Sepulcre, and J. Gozalvez, “Geo-based
scheduling for C-V2X networks,” 2019, arXiv:1906.10370.

[28] T. Sahin and M. Boban, “Radio resource allocation for reliable outof-coverage V2V communications,” in Proc. IEEE 87th Veh. Technol.
Conf. (VTC Spring), 2018, pp. 1–5.

[29] S. Xiaoqin, M. Juanjuan, L. Lei, and Z. Tianchen, “Maximumthroughput sidelink resource allocation for NR-V2X networks
with the energy-efficient CSI Transmission,” IEEE Access, vol. 8,
pp. 73164–73172, 2020.

[30] F. Abbas, G. Liu, P. Fan, Z. Khan, and M. S. Bute, “A vehicle density
based two-stage resource management scheme for 5G-V2X networks,”
in Proc. IEEE 91st Veh. Technol. Conf. (VTC-Spring), 2020, pp. 1–5.

[31] J. Gao, M. R. A. Khandaker, F. Tariq, K.-K. Wong, and R. T. Khan,
“Deep neural network based resource allocation for V2X communications,” in Proc. IEEE 90th Veh. Technol. Conf. (VTC-Fall), 2019,
pp. 1–5.

[32] Z. Ali, S. Lagén, L. Giupponi, and R. Rouil, “3GPP NR V2X mode 2:
Overview, models and system-level evaluation,” IEEE Access, vol. 9,
pp. 89554–89579, 2021.

[33] F. Romeo, C. Campolo, A. Molinaro, and A. O. Berthet, “DENM repetitions to enhance reliability of the autonomous mode in NR V2X
sidelink,” in Proc. IEEE 91st Veh. Technol. Conf. (VTC-Spring), 2020,
pp. 1–5.

[34] Y. Yoon and H. Kim, “A Stochastic reservation scheme for aperiodic
traffic in NR V2X communication,” in Proc. IEEE Wireless Commun.
Netw. Conf. (WCNC), 2021, pp. 1–6.

[35] S. Yi, G. Sun, and X. Wang, “Enhanced resource allocation for 5G V2X
in congested smart intersection,” in Proc. IEEE 92nd Veh. Technol.
Conf. (VTC-Fall), 2020, pp. 1–5.

[36] Z. Ali, S. Lagén, and L. Giupponi, “On the impact of numerology
in NR V2X mode 2 with sensing and no-sensing resource selection,”
2021, arXiv:2106.15303.

[37] C. Campolo, A. Molinaro, F. Romeo, A. Bazzi, and A. O. Berthet, “5G
NR V2X: On the impact of a flexible numerology on the autonomous
sidelink mode,” in Proc. IEEE 2nd 5G World Forum (5GWF), 2019,
pp. 102–107.

[38] H. Ye and G. Ye. Li, “Deep reinforcement learning for resource allocation in V2V communications,” in Proc. IEEE Int. Conf. Commun.
(ICC), 2018, pp. 1–6.

[39] “New WID on NR sidelink enhancement,” LG Electron., Sitges, Spain,
Rep. 3GPP TSG RAN Meeting 86, R1-1903073, Dec. 2019.

[40] L. Cao and H. Yin, “Resource allocation for vehicle platooning in 5G
NR-V2X via deep reinforcement learning,” 2021, arXiv:2101.10424.

[41] H. Ye, G. Ye Li, and B.-H. F. Juang, “Deep reinforcement learning
based resource allocation for V2V communications,” IEEE Trans. Veh.
Technol., vol. 68, no. 4, pp. 3163–3173, Apr. 2019.

[42] M. Zhu, Y. Hou, X. Tao, T. Sui, and L. Gao, “Joint optimal allocation of wireless resource and MEC computation capability in vehicular
network,” in Proc. IEEE Wireless Commun. Netw. Conf. Workshops
(WCNCW), 2020, pp. 1–6.

[43] Y. Hou, C. Wang, M. Zhu, X. Xu, X. Tao, and X. Wu, “Joint
allocation of wireless resource and computing capability in MECenabled vehicular network,” China Commun., vol. 18, no. 6, pp. 64–76,
Jun. 2021.

\end{document}